\documentclass[preprint2]{emulateapj}  
\usepackage{graphicx,multirow}

\shorttitle{STEREO observation of a CME without disk signature}
\shortauthors{E. Robbrecht et al.}

\begin{document}

\title{No trace left behind: \\ STEREO observation of a Coronal Mass Ejection without low coronal signatures}

\author{Eva Robbrecht\altaffilmark{1}, Spiros Patsourakos\altaffilmark{1} and Angelos Vourlidas\altaffilmark{2}}
\altaffiltext{1}{George Mason University, 4400 University Dr., Fairfax, VA 22030, USA} 
\altaffiltext{2}{Naval Research Laboratory, 4555 Overlook Ave SW, Washington, DC 20375, USA}

\email{Eva.Robbrecht.ctr.be@nrl.navy.mil}


\begin{abstract}
The availability of high quality synoptic observations of the EUV and visible corona during the SOHO mission has advanced our understanding of the low corona manifestations of CMEs. The EUV imager/white light coronagraph connection has been proven so powerful, it is routinely assumed that if no EUV signatures are present when a CME is observed by a coronagraph, then the event must originate behind the visible limb. This assumption carries strong implications for space weather forecasting but has not been put to the test. This paper presents the first detailed analysis of a frontside, large-scale CME that has no obvious counterparts in the low corona as observed in EUV and H$\alpha$ wavelengths. The event was observed by the SECCHI instruments onboard the STEREO mission. The COR2A coronagraph observed a slow flux-rope type CME, while an extremely faint partial halo was observed in COR2B.  The event evolved very slowly and is typical of the streamer-blowout CME class. EUVI A 171~\AA\ images show a concave feature above the east limb, relatively stable for about two days before the eruption, when it rises into the coronagraphic fields and develops into the core of the CME. None of the typical low corona signatures of a CME (flaring, EUV dimming, filament eruption, waves) were observed in the EUVI-B images, which we attribute to the unusually large height from which the flux-rope lifted off. This interpretation is supported by the CME mass measurements and estimates of the expected EUV dimming intensity. Only thanks to the availability of the two viewpoints we were able to identify the likely source region. The event originated along a neutral line over the quiet sun. No active regions were present anywhere on the visible (from STEREO B) face of the disk. Leaving no trace behind on the solar disk, this observation shows unambiguously that a CME eruption does not need to have clear on-disk signatures. Also it sheds light on the question of `mystery' geomagnetic storms, storms without clear solar origin (formerly called problem storms). We discuss the implications for space weather monitoring. Preliminary inspection of STEREO data indicates that events like this are not uncommon, particularly during the ongoing period of deep solar minimum. 
\end{abstract}



 

\section{Introduction}

\begin{figure}\centering
\includegraphics[width=\linewidth]{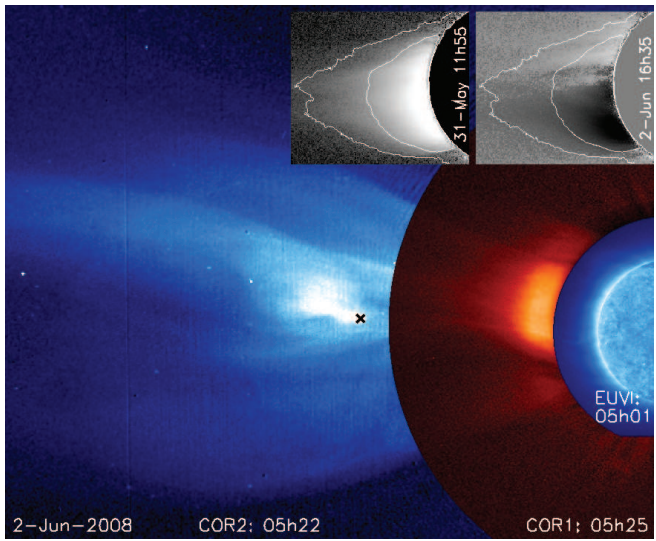}
\caption{CME shown in background removed composite image from the STEREO A spacecraft. The `$\times$' indicates the feature we tracked from EUVI through COR2. The two inserts show the east limb of the Sun in COR1A before and after the eruption (not to scale with the larger image). The first insert (left) shows a background subtracted image, the contours indicate the position of the streamer. The second insert (right) is a difference image with the pre-event image subtracted, showing the disappearance of the southern part of the helmet streamer.  \label{fig:composite}}
\end{figure}

\placefigure{fig:composite}

The relationship of coronal mass ejections (CMEs) to other forms of solar activity has been the subject of numerous studies \citep[see][]{2006SSRv..123..341P}. Nevertheless, no simple causal relation has been found and it seems that no such simple relation exists. CMEs were discovered in the early 1970's with space-borne white light coronagraphs \citep{1973spre.conf..713T,1974JGR....79.4581G}, long after flares and prominences had been observed. It was assumed, then, that the CMEs were simply a product of a flare and/or filament eruption. As CME observations became more common, this causal relationship came under dispute \citep{1993JGR....9818937G,1996SoPh..166..441H} with important implications for Sun-Earth studies. Up to that point, flares and H$\alpha$ filament disappearances had been interpreted as the direct cause of large nonrecurrent geomagnetic storms. We now know that nonrecurrent geomagnetic storms are due to the interaction of the magnetic field of CMEs with the terrestrial magnetic field. While the causal relationship among these forms of solar activity is still under debate, there is a strong belief in the community (including the operational space weather community) that CMEs are strongly intertwined with flares and prominence eruptions to such an extent that one expects to observe either or one of them whenever a CME is observed. For example long duration flares, EUV dimming regions and filament eruptions are routinely used by the NOAA forcasters as proxies for determining the source region and probable propagation direction of a CME. The recent emphasis on the EUV low corona counterparts of CMEs has actually led to the adoption of an EUV full disk imager (but without a coronagraph) for the next generation of operational satellites. But if this is to be a reliable operational strategy, CMEs must always have discernible low corona counterparts. Is this true?

There is enough evidence to believe that CMEs are best correlated with erupting prominences and filaments \citep[][and references therein]{1979SoPh...61..201M,1987SoPh..108..383W,2006SSRv..123...81A}. These comprise quiet sun as well as active region prominences and filaments. However, the reverse is generally not true: filaments can disappear thermally and prominences often erupt in a confined fashion, thus not leading to a CME. On the other hand, flares are more numerous than CMEs. Almost all long duration events (LDE) have an associated CME, because the LDE is caused by the heating of loops due to reconnection below the CME. Again, the reverse is not true. There are large CMEs without flaring and large flares (even X-class) without CMEs \citep[e.g.][]{1994JGR....99.8451F}. SOHO observations have shown  EUV dimmings associated with CMEs \citep[e.g.][]{1998GeoRL..25.2465T}. It is generally believed that the dimming is due to the evacuation of coronal mass and as such it is a reliable proxy for a CME. Dimmings had been observed also in soft X-ray images from Skylab \citep{1976SoPh...48..381R} and Yohkoh \citep{1997ApJ...491L..55S}. \cite{2008A&A...478..897B} confirmed the close relationship between CMEs and CDS dimmings, but again no one-to-one correspondence was found; only up to 84\% of the CMEs could be traced back to a CDS dimming.

For the majority of CMEs, especially fast ones, it is generally easy to identify a number of associated low coronal and chromospheric signatures. But there exist several examples of white light halo CMEs with no such association, even though the in situ data suggests arrival at Earth. Using a comprehensive set of data, \cite{2003ApJ...582..520Z,2007JGRA..11210102Z} and \cite{2005AnGeo..23.1033S} searched for source regions of geomagnetic storms (respectively with Dst $\le -100$ nT and Dst $\le -50$ nT). The identification process itself did not seem straightforward: only half of the geomagnetic storms had a clear association with a unique disk-signature and CME.  The remainder of the storms had multiple candidate sources or no candidate. 11\% of the storms studied by \cite{2007JGRA..11210102Z} could not be linked to a signature on the disk, but they were all caused by slow partial halo CMEs. \cite{2005AnGeo..23.1033S}  reported that about 20\% of the geoeffective ICMEs were not preceded by an identifiable frontside halo CME (they used SOHO/EIT data to separate backsided CMEs). The standard explanation has always been that halo CMEs lacking on-disk signatures must be backsided and were catalogued as such. While this is correct for some events, there have been cases of geomagnetic storms associated with apparently backsided CMEs \citep[e.g. in ][]{2005AnGeo..23.1033S}. These associations have always been controversial because it is hard to imagine how a CME directed away from the Sun-Earth line could have any geoeffective potential. Appropriately, these geomagnetic storms are called `problem storms'. An early example is the January 6-10 1997 event \citep{1998GeoRL..25.2469W} for which the corresponding white light halo CME was only identified post-facto. Two other early examples of problem storms were the April 22-23 1997 and the June 9 1997 storms that reached a Dst of -107 nT and -84 nT, respectively. These storms were driven by magnetic clouds with flux-rope characteristics, but could not be associated with any frontside (halo) CME \citep{2000JGR...105.7491W}. The problem would disappear if the origins of these storms have been misidentified simply due to the lack of observable EUV or other low coronal signatures. The implications for space weather studies are obvious.

In this paper, we analyze STEREO/SECCHI observations of a streamer blowout CME without a clear source region other than the quiet sun.  Thanks to the wide angle separation (53$^{\circ}$) of the STEREO spacecraft we were able to study the CME and its source region edge-on in STEREO A and face-on in STEREO B (\S~\ref{sec_observations}).  We analyze the kinematics and the mass budget. The event was captured in situ by the STEREO B spacecraft where a magnetic cloud (MC) was observed. If a MC with similar plasma parameters and magnetic field strength but southward Bz impacted the Earth, it may have caused a geomagnetic disturbance at
a moderate level with minimum Dst $\approx -70$ nT (Yan Li, private communication). This event is thus a good example of a `problem storm'. In STEREO B only an extremely faint halo CME was observed lacking any obvious disk counterparts. We suggest that the CME originated from high in the corona and therefore caused no observable dimming. We conclude in \S~\ref{sec_concl} discussing the implications for CME science and space weather operations. These events are barely observable from the Earth's vantage point and are thus unpredictable for their geoeffective potential.

\begin{figure*}\centering
\includegraphics{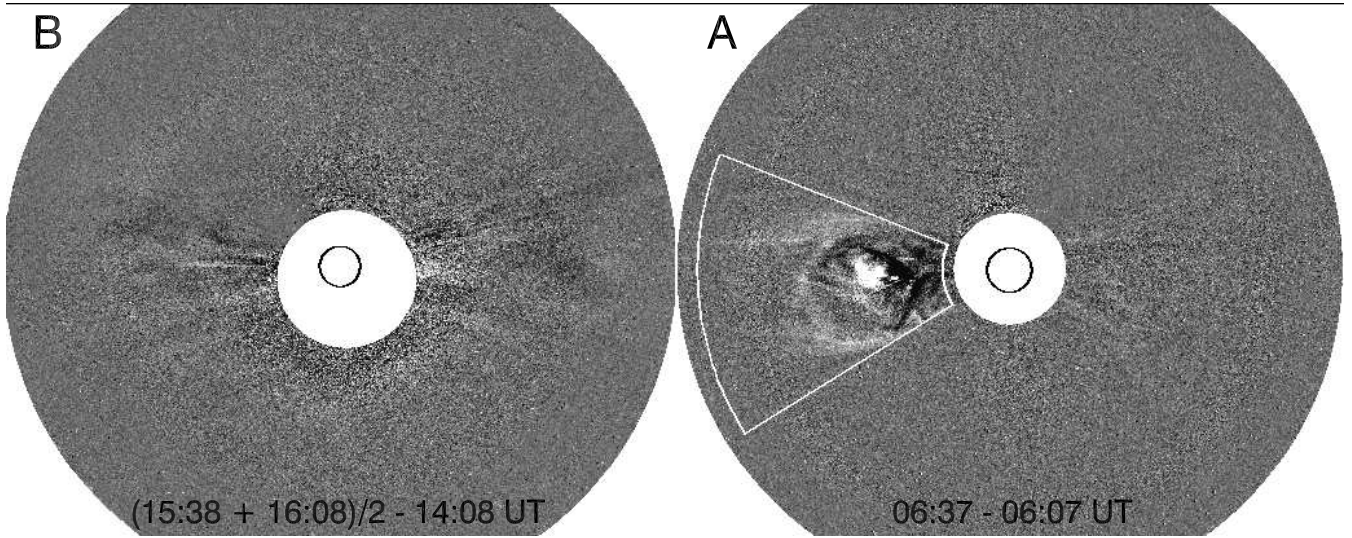}
\caption{SECCHI/COR2 observations of the June 2, 2008 CME. The images are running differences of total brightness images. Only a very faint halo was observed in COR2B (better seen in the movie). To enhance the contrast in the COR2B image we averaged two images before subtracting a previous image.  The times at which the original images were taken are printed at the bottom in each figure.The black circles mark the position and size of the Sun. The white sector in the COR2A image indicates the region used to calculate the CME mass (see \S~\ref{mass_origin}). At the time the COR2A image was taken, the halo CME was not yet visible in the COR2B FOV, so the times are different in the two frames. \label{fig:cor2a-b}}
\end{figure*}

\placefigure{fig:cor2a-b}

\begin{figure*}\centering
\includegraphics[width=.49\linewidth]{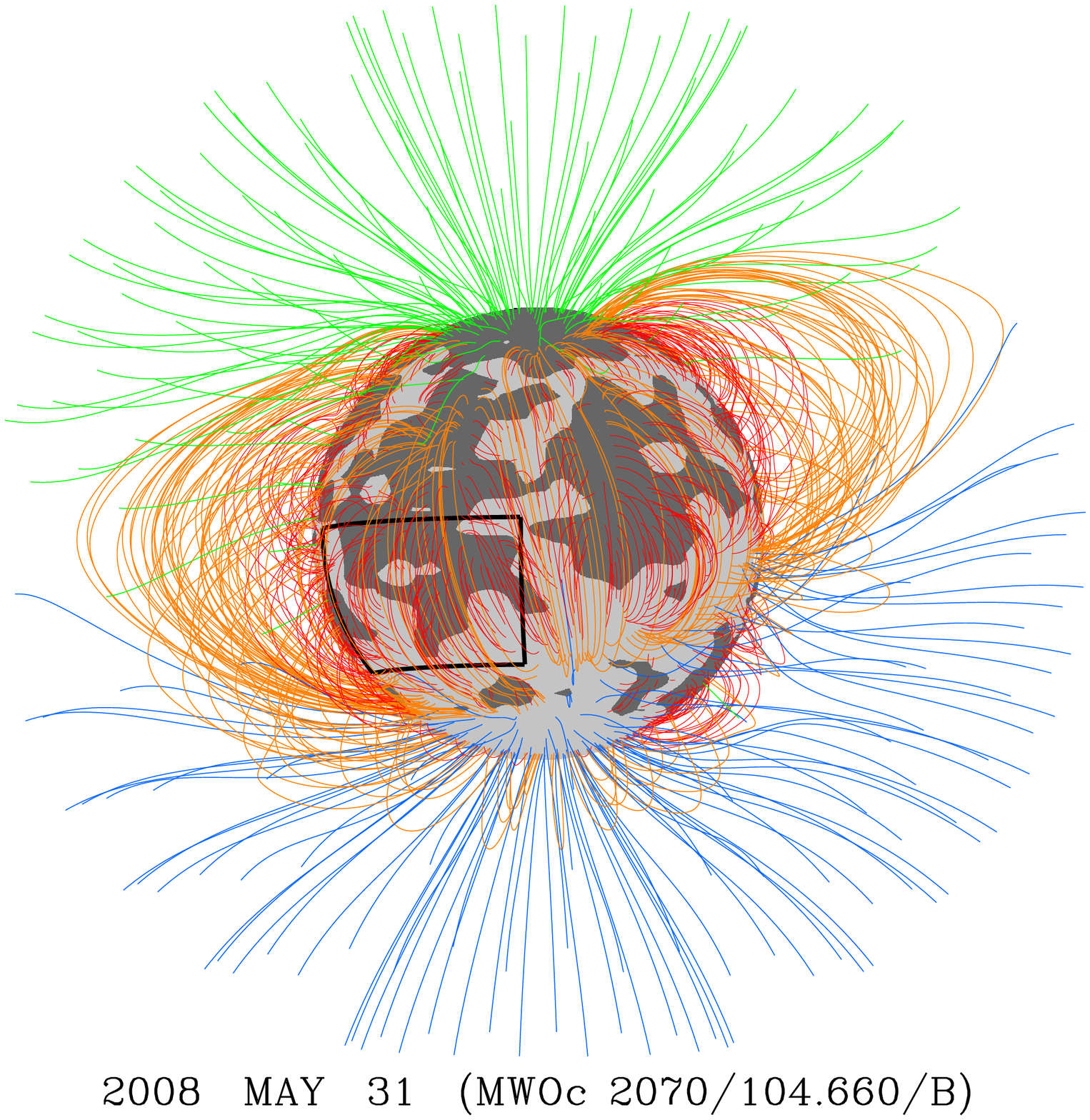}
\includegraphics[width=.49\linewidth]{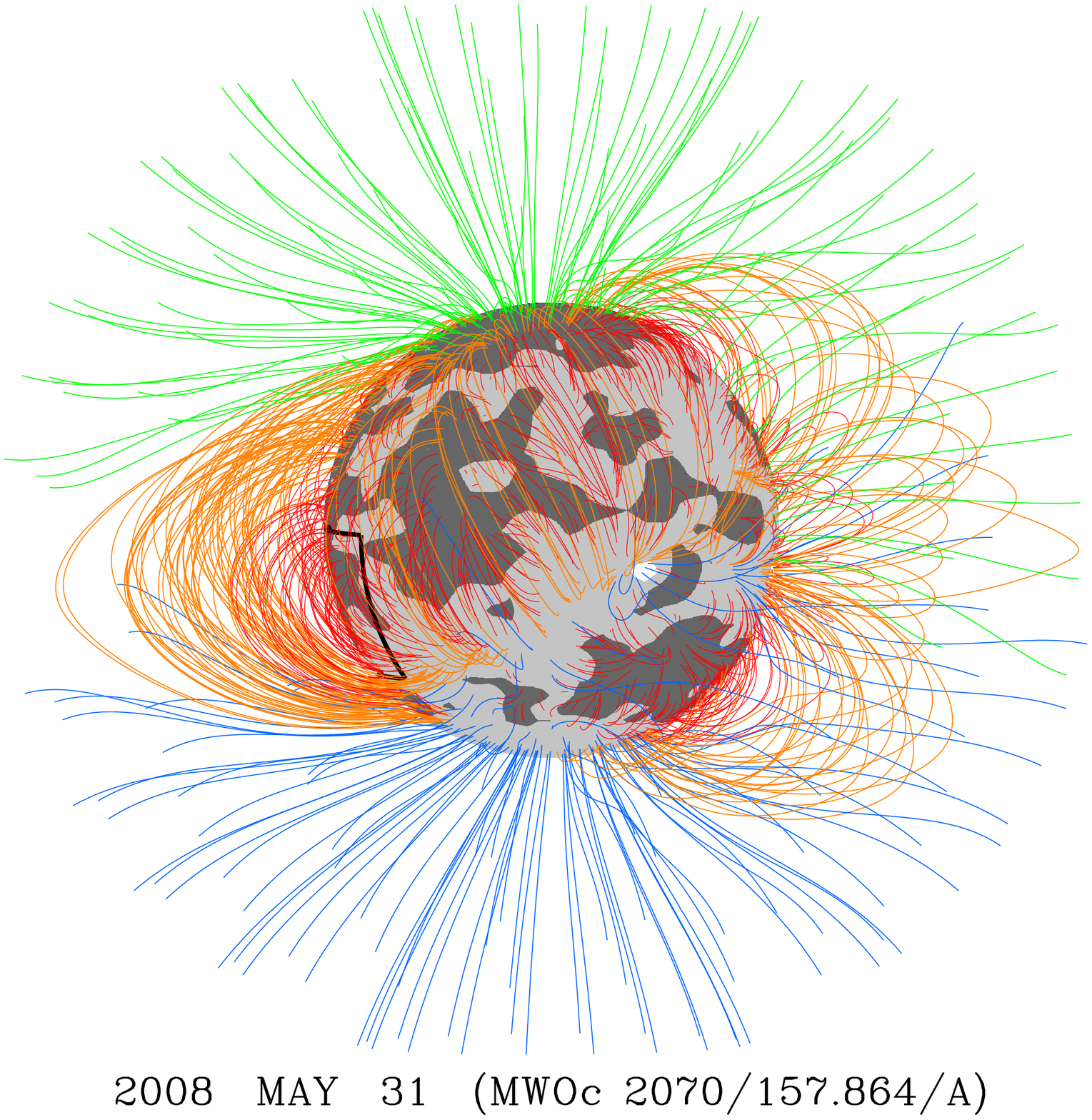}
\caption{Potential Field Source Surface magnetic field extrapolations based on MWO data. The extrapolations are centered as viewed from STEREO B (left) and STEREO A (right) on May 31, 2008 00:00 UT. The black box marks the probable source region, it is the same box as the one in Figure~\ref{fig:context}. Green (blue) refers to open field lines of negative (positive) polarity, and orange (red) refers to long (short) closed field lines: the red field lines reach heights up to 1.5 R$_\odot$, the yellow field lines between 1.5 R$_\odot$ and 2.5 R$_\odot$. At the photosphere, light grey areas show positive magnetic flux ($0 < B_r < 10$ G), and dark grey areas show negative flux ($-10 < B_r < 0 $ G).  (Image credit: Yi-Ming Wang)
\label{fig:pfss}}
\end{figure*}

\placefigure{fig:pfss}

\placetable{table:instruments}
\begin{deluxetable*}{c c c c c c c}
\tablecaption{STEREO/SECCHI instrument details (References are in text)
\label{table:instruments}}
\tablewidth{0pt}
\tablehead{\colhead{{\bf Instrument}} &\multicolumn{4}{c}{EUVI}  &  \colhead{COR1}  &  \colhead{COR2}  } 
\tablecolumns{7}
\startdata 
{\bf Type} &   \multicolumn{4}{c}{EUV Telescope}& Coronagraph &  Coronagraph \\ 
{\bf Bandpass} & 171 \AA& 195 \AA& 284 \AA& 304 \AA & White-Light (650-660 nm)& White-Light (650-750 nm)\\
{\bf Cadence} & 2.5 min& 10 min& 20 min & 10 min& 8 min & 15 min\\ 
{\bf FOV from sun center} &\multicolumn{4}{c}{$0 - 1.7 \, R_{\odot}$ } & $1.4-4 \, R_{\odot}$ & $2.5-15 \, R_{\odot}$  \\ 
{\bf Pixel Size} &  \multicolumn{4}{c}{1.6 arcsec} & 7.5 arcsec & 15 arcsec 
\enddata
\end{deluxetable*}

\section{STEREO Observation and measurements}\label{sec_observations}
Our analysis is based on data from the Extreme Ultra-Violet Imaging Telescope (EUVI) and white-light images from the COR1 and COR2 coronagraphs onboard SECCHI \citep{2008SSRv..136...67H}. Details of the instruments are given in Table~\ref{table:instruments}. The COR1 and COR2 images are total brightness images. Figure~\ref{fig:composite} is a composite image of the SECCHI A observations (a composite movie is available online). Following a streamer swelling that lasted for about two days, the CME entered the COR2 field of view on June 2, 2008. Figure~\ref{fig:cor2a-b} shows a running difference snapshot of the event taken in the two COR2 telescopes (as in all the Figures in this paper, the STEREO A perspective is on the right). The CME has a classical flux-rope morphology \citep{2000ApJ...534..456V} in COR2A and is an extremely faint halo in COR2B. Halo CMEs are faint because the Thomson scattering, which forms the white-light image, is most effective in the plane of the sky.  Therefore Figure~\ref{fig:cor2a-b} immediately suggests that the CME erupted close to the plane-of-sky (POS) of STEREO A. 

The CME erupted from below a helmet streamer at the east limb, as seen by the STEREO A coronagraphs. The two inserts in Figure~\ref{fig:composite} show the initial streamer and its partial disappearance after the CME erupted. The streamer swelling, its disappearance, slow evolution and flux-rope structure clearly identify this event as a streamer-blowout CME \citep{2002stma.conf..201V}.  The {\it only partial} disappearance of the streamer can give us a clue as to the origin of the CME with the help of the potential field source surface (PFSS) magnetic field extrapolation shown in Figure~\ref{fig:pfss}. The extrapolation method is described in \cite{1992ApJ...392..310W,1995ApJ...447L.143W}. The preexisting helmet streamer can be readily identified with the overlying closed (red and orange) field lines. The part that erupted overlies the red field lines that close below a height of 1.5\,R$_\odot$. These field lines only occupy the lower latitudes of the streamer.

Figure~\ref{fig:context} shows the full disk corona in 171~\AA\, prior to the eruption. This is a typical solar minimum corona lacking big active regions, and is dominated by the small scales characteristic for the quiet sun. The arrow indicates the bright structure that traveled outward and eventually developed into the CME core. The morphology of the feature (concave shape, bright core) suggests that it is the bottom of a flux-rope, so large that the top is outside the field of view of the EUVI image. We draw a parallel with the cavity/bright rim that is observed in the north west (top right) corner of the EUVI B image. To compare the sizes, we measured the widths of both cavities, similarly to \cite{2006ApJ...641..590G}. The cavities were measured in the EUVI 171  \AA\, images at 1.15\,R$_{\odot}$ from sun center and resulted in widths of around 25$^\circ$ and 11.5$^\circ$, for the A and B cavity, respectively. We thus estimate that the cavity of interest here is a factor $\sim2.2$ wider than the one observed in B.  \cite{2006ApJ...641..590G} found a largely constant aspect ratio of cavity width to height, hence we can deduce that our flux-rope/cavity is about 2.2 times higher than the one observed in EUVI B. This large height will be of relevance in \S~\ref{sec_disc} where we discuss the lack of a dimming.

\subsection{Estimation of the true CME direction of propagation}\label{subsec:direction}

\begin{figure*}\centering
\includegraphics[width=\linewidth]{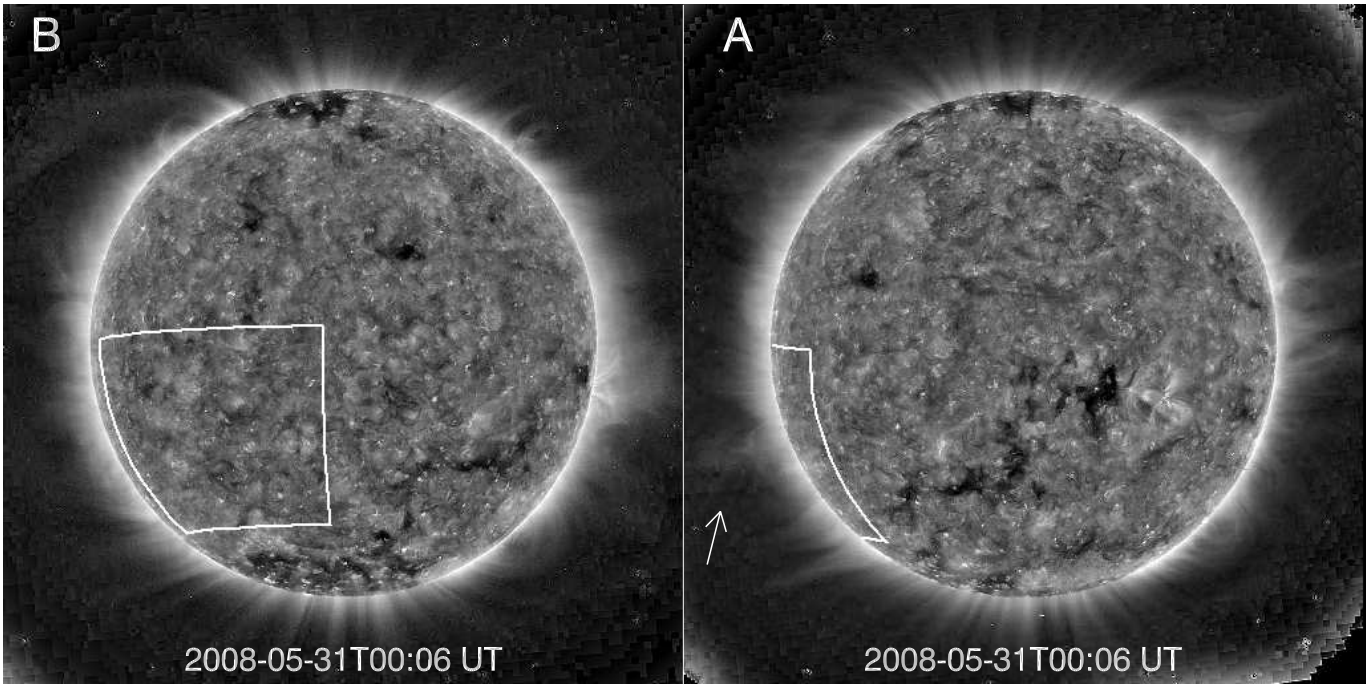}
\caption{Simultaneous images taken by EUVI A and B showing the corona in the 171 \AA\, line before the start of the eruption. The field-of-view in both images is cropped at 1.35\,R$_{\odot}$ (from suncenter). The white box marks the probable source region of the CME, it extends from $30^{\circ}$ to $100^{\circ}$ in Carrington longitude and from $-40^{\circ}$ to $0^{\circ}$ in latitude. The arrow indicates a bright structure, which is the bottom part of the erupting flux-rope. The images were contrast enhanced using a wavelet algorithm \citep[adapted from][]{2008ApJ...674.1201S}.
\label{fig:context}}
\end{figure*}

\placefigure{fig:context}

\begin{figure}\centering
\includegraphics[width=\linewidth]{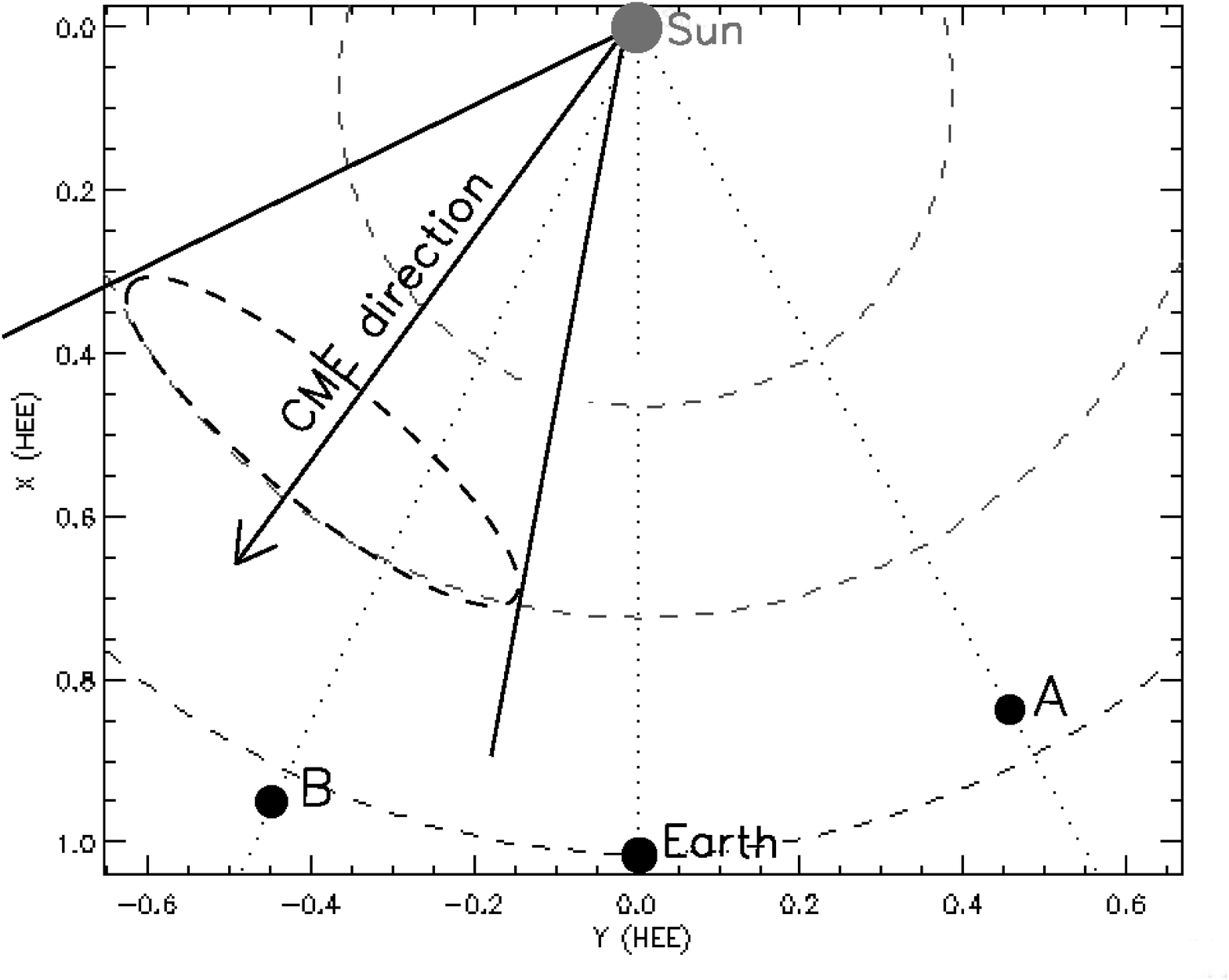}
\caption{Schematic view of the CME direction projected on the ecliptic plane (top view). We assume a circular cone model with an opening angle equal to the CME width as measured from STEREO A (54 degrees). The separation angle between the A and B spacecraft was 53 degrees. \label{fig:direction}}
\end{figure}

At the time of the eruption, the separation angle between the STEREO A and B spacecraft was 53 degrees. This separation allowed us to observe the CME simultaneously edge on (in A) and face on (in B). Figure~\ref{fig:direction} shows a schematic representation of the position of the CME relative to the STEREO spacecraft.  We used several methods to derive the true propagation direction of the CME:
\begin{itemize}
\item The observation of a bright CME in the COR2A and an extremely faint halo in the COR2B images (see Figure~\ref{fig:cor2a-b}) means that the CME propagated towards STEREO B and lies close to the POS of STEREO A. 
 \item Polarization analysis of COR2 polarized brightness (pB) images suggests that the CME lay in the POS of STEREO A $\pm 20^{\circ}$. 
\item Applying a forward modeling technique to the CME, \cite{2009SoPh..256..111T} estimate an angle of $26\pm10^{\circ}$ out of the POS of STEREO A (frontsided). 
\item The in situ observation of the arrival of a MC on June 6 in STEREO B (Yan Li, private communication) confirms that the CME propagation path had a large component in the STEREO B direction. 
\end{itemize}
From the above, we estimate that the CME propagated at approximately $40^{\circ}$ east from the Sun-Earth line. At the time of maximum acceleration (around 20:00 UT on 1 June 2008) this direction corresponds to a Carrington longitude of $65^{\circ}$. This also gives us an initial estimate of the position of the source region. The derived Carrington longitude corresponds to the position of the east limb as seen from STEREO A at 00:00 UT on 31 May 2008 (see Figure~\ref{fig:context}) and suggests indeed that this CME erupted from the pre-existing flux-rope indicated with the white arrow.

\subsection{CME kinematics derived from STEREO A}

During its evolution, the CME front is barely visible in COR1, but it can be clearly seen when it enters the COR2 field-of-view (FOV) around 19:00 UT on 1 June 2008. The evolution of the event in the EUVI-COR1-COR2 combined  FOV spans more than three days starting on 31 May  2008. The slow evolution of the event has allowed us to measure the kinematics of the event in great detail (Figure~\ref{fig:speed}, a-b). Because of the lack of a front in the COR1 images, we traced the back of the CME core, which is the best obsered feature in all three instruments. It is indicated with an `$\times$'  in Figure~\ref{fig:composite}. Because of the CME expansion the core travels slower than the leading edge. The same feature can be seen off-limb in EUVI A (Figure~\ref{fig:euvi}) as it leaves the Sun. We only show the 171 \AA\, component where it is best observed. It is also visible in 195 \AA, but not in the other 2 channels. We can identify three stages in the CME evolution: (1) a slow rise phase that starts around 20:36 UT on 31 May  2008, (2) a quasi-constant acceleration phase starting around 21:00 UT on 1 June 2008 and (3) a constant velocity phase. The asymptotic velocity of around 300 km/s is only reached at around 20\,R$_{\odot}$ in the STEREO/HI (Heliospheric Imager) FOV. Here we focus on the first two phases.  

Several hours before the actual eruption, EUVI A images show off-limb a helical structure rolling around its axis (see top panel in Figure~\ref{fig:euvi} and online movie). This activity may be interpreted as flux-rope activation or formation. Then a concave structure below the helical structure starts to rise very slowly in the EUVI FOV at around 10 km/s (shown in the next two panels in Figure~\ref{fig:euvi}). This quasi-steady phase can be best approximated by a linear fit, until the feature reaches $\sim2$\,R$_{\odot}$. During this phase, the white light streamer brightens and swells. Once in the coronagraph FOV, the feature starts to accelerate gradually until it reaches a velocity of around 200 km/s at 13\,R$_{\odot}$. At this stage it has developed into the bright core of the CME. We find that this acceleration phase is well described by a second order curve (red curve in Figure~\ref{fig:speed}, a-b). 

Under certain conditions different initiation models predict different height-time profiles \citep[e.g.][]{2008ApJ...674..586S}. Motivated by this, we also fitted an exponential function (green, dash-dotted) and a power law function (blue, dotted). Up to now, and to the best of our knowledge, researchers have fit only second order curves to these gradual type events. From Figure~\ref{fig:speed} (a-b) we can see that the three colored curves describe relatively well the acceleration phase. This implies that, at least for this event, it is impossible to distinguish between the existing models based on the kinematics alone. 

\subsection{Localization of the source region in EUVI B}\label{subsec:stereob}

From our estimation of the true propagation direction (\S \ref{subsec:direction}), we derive that EUVI B observes the source region face-on.  Figure \ref{fig:context} shows a pair of images taken simultaneously by EUVI 171 \AA\, before the eruption. The white boxes extend from $30^{\circ}$ to $100^{\circ}$ in Carrington longitude and from $-40^{\circ}$ to $0^{\circ}$ in latitude. The latitudinal extent of the box is derived from the edge-on view seen in EUVI A. In longitudinal direction, the box is centered at $65^{\circ}$, which is the direction of the CME derived in \S~\ref{subsec:direction} roughly at the time when the CME detaches from the sun. Close inspection of EUVI B data (all wavelengths) of the source region does not reveal any clear on-disk signature in any of the four EUVI wavelengths. There is no active region or sunspot group, no apparent filament eruption that caused the CME, no GOES X-ray emission (see Figure \ref{fig:speed}c), no post-eruption loop arcades, no EUV wave and no obvious EUV dimming. We strongly encourage the reader to inspect the movie that is available online (the movie runs from May 31st to 2 June 2nd). Figure \ref{fig:euvi} shows five snapshots from EUVI A and B 171 \AA\, illustrating the low-coronal evolution during the event. As can be better seen in the accompanying movie, we only detect small scale activity in the B images, which is ubiquitous in the quiet sun. Several small features are visible that may be related to the eruption. For example we encircled in yellow (f3) a small bright patch that could be evidence of small-scale heating. Also, the small prominence indicated in orange (f2) that was seen off-limb in EUVI A 304 \AA\, cannot be readily identified in any of the EUVI B channels, which suggests that it is not very dense.  The footpoint of it in the EUVI B FOV is indicated with a diamond. It is also interesting to note that we observed a small filament in H$\alpha$ just after the CME erupted. Its shape is outlined in turquoise in the bottom frames of Figure \ref{fig:euvi}. The filament forms fast, is not very conspicuous and disappears quickly between 21:00 and 22:00 UT (S. Martin, private communication). While this filament is not related to the observed CME, it indicates the presence of a filament channel at the time of the eruption \citep[e.g.][]{1998SoPh..182..107M}. Further, a small dimming area is indicated (magenta box in f5).  While these small features may have a relation to the observed CME, their scale is much smaller than the CME itself. Moreover, the movie illustrates that similar activity is observed outside the box, thus it is difficult to decide which features are connected to the observed CME.

The most surprising result is the lack of observation of an eruptive filament in the EUVI B images. We expected to see something given the clear helical structure seen in both EUVI A and COR2A. To make sure, we examined H$\alpha$ images from varied sources, but no H$\alpha$ filament was observed prior to the CME.  H$\alpha$ images are only available from Earth, which solar view is between that of STEREO A and B. On May 31, 2008 00:00 UT, the east limb of the sun, as seen from the Earth, has Carrington longitude $39.5^{\circ}$, which is $25.5^{\circ}$ east of $65^{\circ}$, our initial estimate for the source region location from \S~\ref{subsec:direction}. If a filament had erupted from around that longitude, we should have observed it in the H$\alpha$ images. 

\subsection{CME mass measurement}\label{mass_origin}

\begin{figure}\centering
\includegraphics[width=\linewidth]{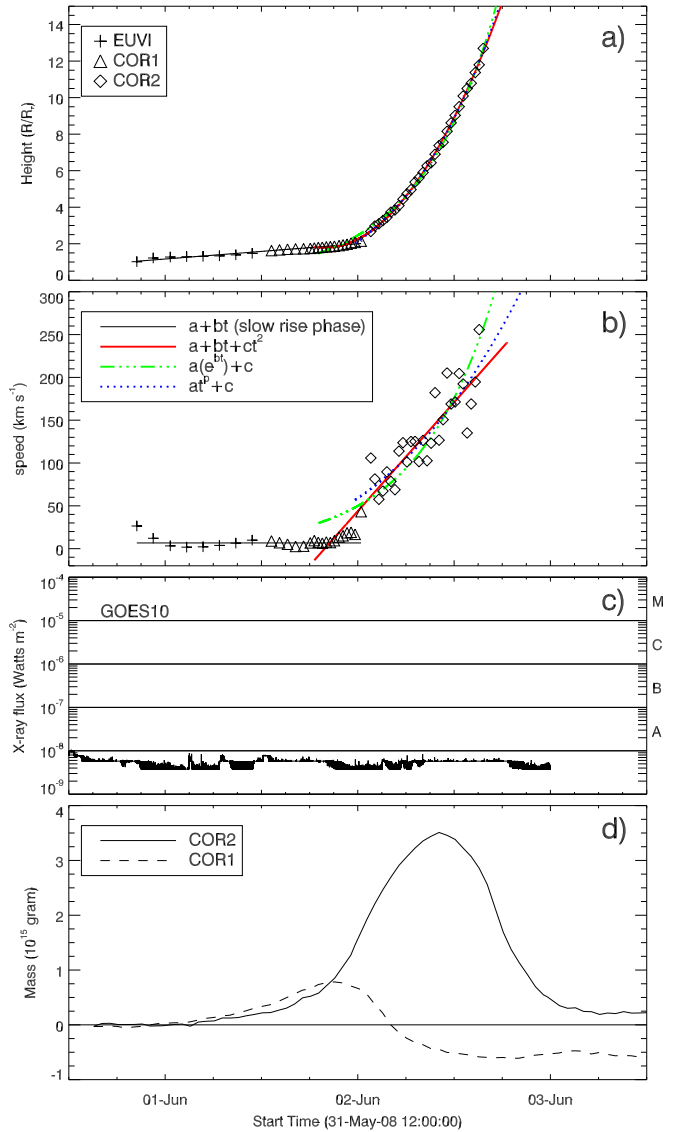}
\caption{(a) Height (from sun center) and (b) speed profile of the CME.  The feature indicated with `x' in Figure \ref{fig:composite} is traced from EUVI-171 to COR2. Three functions (second order, exponential and power law) are fitted to the acceleration phase. The initial slow rise is fitted with a constant speed profile. (c) The GOES X-ray flux shows no flaring activity during the period of interest. (d) CME mass as a function of time through the STEREO A coronagraphs FOV. The mass is calculated by integrating the electron density in a fixed sector in base-difference images (illustrated in Figure~\ref{fig:cor2a-b}). When the CME leaves the COR1 FOV (dashed line) a mass depletion (`negative mass') can be observed indicating a partial streamer blowout. \label{fig:speed}}
\end{figure}

\placefigure{fig:speed}

\begin{figure}\centering
\includegraphics[width=\linewidth]{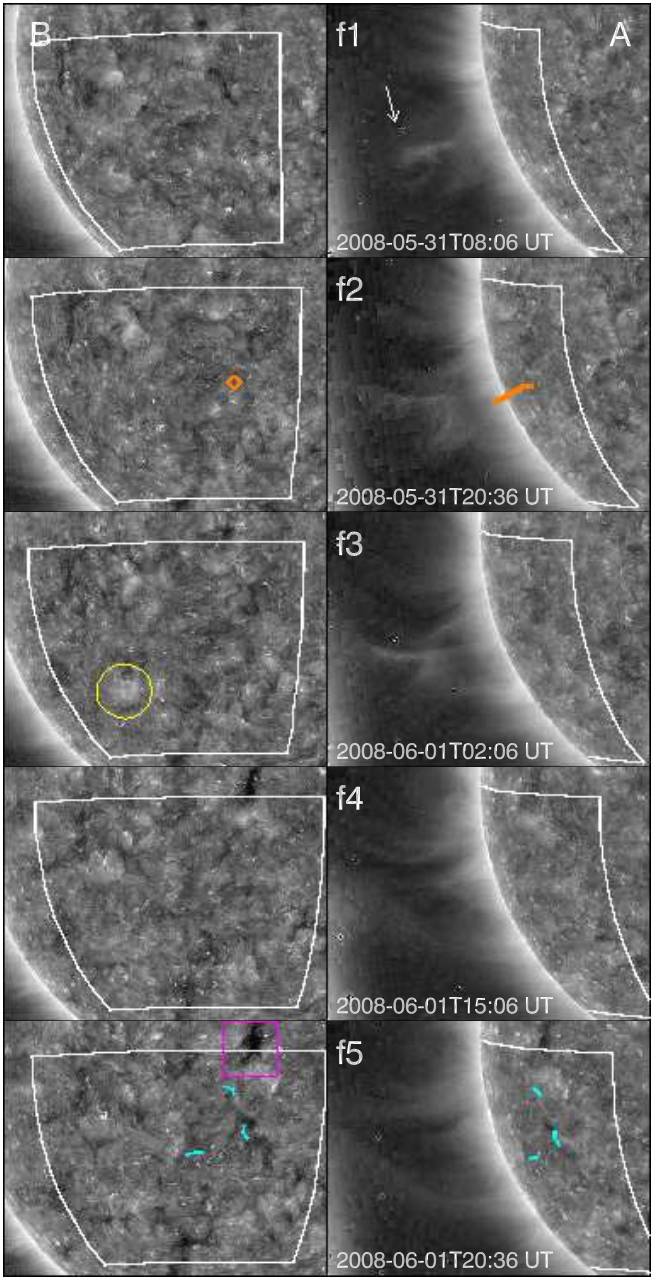}
\caption{Five snapshots from EUVI 171 \AA\, illustrate the low-coronal evolution during the event (A is on the right, B is on the left, time runs from top to bottom). The images were contrast enhanced using a wavelet technique \citep[based on][]{2008ApJ...674.1201S}. Several stages are seen in EUVI A: helical motion in f1 (white arrow), rising flux-rope in f2 and f3 and detaching flux-rope f4 and f5. Movies are made available on the website. The white box is the same as in Figure~\ref{fig:context}. Several small features are indicated in color: orange in f2: a faint filament that was observed in EUVI A in 304 \AA\, (the footpoint is indicated in B), yellow in f3: a small signature of heating, possibly as part of the eruption, magenta in f5: small dimming or appearance of coronal hole and finally turquoise in f5: a small filament appeared and disappeared quickly in H$\alpha$. \label{fig:euvi}}
\end{figure}

\placefigure{fig:euvi}

Because of the detailed coronagraph observations in STEREO A, we can assess the height from which the CME material originates by measuring the mass flux in the COR1 and COR2 fields of view (shown in Figure ~\ref{fig:speed} (d)). The CME mass is estimated by integrating over a region of interest (ROI), the excess brightness of a given image relative to a pre-event image. We defined the ROI as a sector bounded by the CME edges in position angle and by 1.8 - 3.7 R$_{\odot}$ in COR1 and 3 - 14 R$_{\odot}$ in COR2 in radial direction (the COR2 sector is drawn in Figure\ref{fig:cor2a-b}), thus the curves in Figure \ref{fig:speed} (d) represent the mass evolution in these ROIs. The maximum mass measurement in the COR2 FOV is a representative number for the total CME mass, being $\sim 3.5 \times 10^{15}$ g which is normal for such events \citep[e.g.][]{2002stma.conf..201V}. The COR1 FOV is too small (in radial direction) to capture the whole CME at one instant of time, therefore the COR1 curve does not reach the full CME mass.  To obtain a rough cross-calibration between the two instruments, we compared the mass measurement in a common sector (from 2.4 to 3.7 R$_{\odot}$) while the CME traveled through this ROI. The COR1 mass curve was  on average 20\% lower than the COR2 mass curve. 

The streamer swelling is represented by the initial slow rise in the mass curves.  When the CME enters the COR2 FOV a sharper increase can be seen, until the leading edge of the CME reaches the outer edge of the COR2 ROI after which the mass decreases back to its pre-event level. As we can see from the dashed curve, the corona in the COR1 FOV does not recover its pre-event level showing an apparent `negative mass'. This indicates that the COR1 corona was depleted. As a check, we did the same exercise for the 26 April 2008 CME, which was not a streamer blowout and had a clear source region and dimming. The COR1 mass curve for that event did return back to its pre-event level, implying that most of the CME mass originated from below the COR1 FOV. Thus, we can say with confidence that a large fraction of the CME mass in our event originated from the COR1 FOV. 

Standard assumptions in mass measurements are (1) that all of the CME material is located in the plane of the sky (therefore, it gives a lower limit for the true CME mass) and (2) that the corona is completely ionized \citep{2000ApJ...534..456V}. Since our CME lies close to the plane of the sky of STEREO A (where the mass measurements are made) these measurements are close to the real values. 

\subsection{Source region size estimation}\label{sec:source}

Arguably, the most intriguing aspect of our observations is the lack of a well-defined source region of the event in EUVI B data (see \S~\ref{subsec:stereob}). Using the information from the two viewpoints, we draw some estimates about the size of the source region along $L_{\mbox{\tiny{SR}}}$ and across $\theta_{\mbox{\tiny{SR}}}$ the neutral line. Often the length of the erupting filament is used as a proxy for the length of the neutral line. Based on a large statistical survey of observations of erupting filaments and their associated CMEs, \cite{2004A&A...422..307C} found that the observed lengths of the filaments ranged from a few degrees up to more than 40$^\circ$ and show a Poisson-type distribution peaking in the range 6$^\circ$ - 12$^\circ$. We do not directly observe a filament and therefore cannnot measure its length, but from the edge-on orientation of the CME and the PFSS magnetic field extrapolation we estimate that the orientation of the neutral line in question is more or less in the longitudinal direction. Assuming that the two small features indicated in Figure~\ref{fig:euvi} were both related to the CME, their longitudinal separation gives us a lower limit of $L_{\mbox{\tiny{SR}}}\approx35^\circ$.  A second estimation can be made using the 3D information derived from a forward modeling technique \citep{2009SoPh..256..111T}.  This technique models the CME as a hollow flux-tube (see schematic view in their Figure 1). The separation angle $2\alpha$ between the legs of the flux-tube and the aspect ratio $\kappa$ of the tube give an upper limit for the length of the source region. From the values given by \cite{2009SoPh..256..111T} for our source region (see their Table 1b), we find then $L_{\mbox{\tiny{SR}}}\le 2\alpha+\arctan(\kappa)\approx36\pm 10^\circ$. 

Another interesting quantity is the size $\theta_{\mbox{\tiny{SR}}}$ of the source region across the neutral line, which should be commensurate to the footpoint-separation of the associated erupting arcade. \cite{2007ApJ...668.1221M} determined a scaling-law relating the CME width $\theta_{\mbox{\tiny{CME}}}$ (which is what we observe edge-on in EUVI A) to the size of the associated erupting arcade. $\theta_{\mbox{\tiny{CME}}}$ corresponds to the CME angular span when the CME is  fully developed in the COR2 FOV and has reached lateral pressure balance with the surrounding magnetic fields. This is thought to occur in the outer corona (i.e $ > $2\, R$_{\odot} $). The scaling-law \citep[e.g. Equation 20 in][]{2007ApJ...668.1221M} can be written as
\begin{equation}
\theta_{\mbox{\tiny{SR}}} \approx \sqrt{\frac{1.4}{<B_r>}} \theta_{\mbox{\tiny{CME}}} \,\cdot
\end{equation}
 $<B_r>$ is the absolute radial magnetic field averaged over the extended quiet sun area enclosed within the small box of Figure~\ref{fig:context}. A full disk photospheric magnetogram from SOLIS \citep{2003SPIE.4853..194K} was used for this calculation (registered from 20h55 to 21h42 on June 1st 2008).  We found $<B_r>\, \approx 3$ Gauss, which indicates a very weak field in the region of interest. Similar values were obtained from a Mt Wilson magnetogram (Yi-Ming Wang, private communication).  Further, we use the CME angular span $\theta_{\mbox{\tiny{CME}}}\approx 54^\circ$ as taken from the online CACTus catalog \citep{2004A&A...425.1097R}. Using these values, we find a source region size $\theta_{\mbox{\tiny{SR}}} \approx 34^\circ$. This value is comparable to the latitudinal width of the box we derived at first sight from the edge-on view in EUVI A in Figure~\ref{fig:context}. This width is commensurate with the footpoint separation of the red magnetic field lines (see Figure~\ref{fig:pfss}) that erupted. 

Summarizing the above, we deduced a source region size of about $36^\circ\times34^\circ$. These values of the source region should be viewed only as estimates. They nevertheless point to a rather extensive source area.  The main conclusion is that we expect a large source region, something which is in concert with the inferences of a large flux-rope mentioned in \S~\ref{sec_observations}. More refined calculations could pin down further the dimensions of the elusive source region. For example, we plan to determine the detailed energetics of our event as well as the magnetic properties (e.g. fluxes) of the associated magnetic cloud. These values would allow some further estimates of the size of the source region, to be deduced from the conservation of the total CME energy and of the magnetic flux in magnetic clouds respectively.

\section{Discussion}\label{sec_disc}

The analysis in \S~\ref{subsec:stereob} shows that the pre-eruption site is right in the field-of-view of EUVI B. But how can we explain the lack of activity which prevents us from identifying the site? First, let us consider the three forms of solar activity that are most often associated with CMEs: flaring, EUV dimming and, prominence/filament eruption. 

\subsection{Flaring hypothesis}
Flaring is usually expected for CMEs associated with active regions. As is obvious from Figure~\ref{fig:euvi}, no active region was present anywhere on the disk as seen from EUVI B. The CME originated in the quiet sun and the lack of flaring is not surprising. Besides energetic flares, polar crown filament eruptions are often accompanied by long post eruptive arcades seen in EUV and soft X-rays. This type of flaring was not observed for our event, either on-disk \citep[e.g. as in][]{1995JGR...100.3473H} or off-limb \citep[e.g. as in][]{2007ApJ...671..926S} which could imply that the field is closing at a large height (small densities) and that the energy release was small resulting in weak heating unobservable by EUVI.

\subsection{Prominence hypothesis}
The kinematics of our CME and its morphology characterize it as a gradual event usually linked to a prominence eruption. Why then is there no clear prominence visible in the EUVI B images, nor in H$\alpha$? EUVI A images provide an important clue: they show a very high flux-rope (as derived in \S~\ref{sec_observations}). The combined observation of a flux-rope, only visible off-disk in the hotter EUVI channels and a small and tenuous filament, barely discernible in the EUVI A 304~\AA\, implies a hot flux-rope with small amounts of cool material. 

The presence of a cavity and the PFSS extrapolation confirm that the CME originated from a polarity inversion line and shows that the existence of an H$\alpha$ filament is not a necessary condition for the eruption. Filament channels coincide with polarity inversion lines and when chromospheric mass loads into the channel, a filament is formed.  As most CME initiation theories, models and observations suggest, the important agent for an eruption is the existence of a filament channel only, not the chromospheric mass. For example, \cite{2004SoPh..219..169L} pointed out that the magnetic configuration of the filament channel is more important than any mass loading for CME initiation. However, there is a widespread belief that H$\alpha$ filament disappearances are reliable proxies for CMEs. We would like to caution observers and space weather forecasters against an over reliance on H$\alpha$ data for disk signatures of eruptions. Our observation shows that CMEs can erupt without having a mass-loaded (and thus observable) filament. 

\subsection{Dimming hypothesis}\label{sec:dimming}

\begin{figure}\centering
\includegraphics[width=\linewidth]{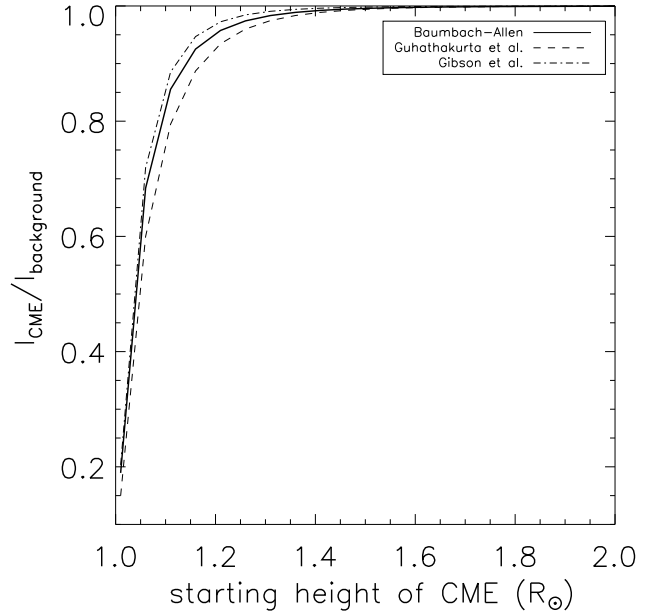}
\caption{Calculation of the EUV dimming due to coronal material removed by a hypothetical CME. The curve shows the ratio of EUV intensities $I_{\mbox{\tiny{CME}}}/I_{\mbox{\tiny{BG}}}$ as a function of the starting height of the CME (measured from sun center). A ratio below one indicates an observable dimming. As can be seen, no dimming is expected for CMEs originating from above 1.4\,R$_{\odot}$. 
\label{fig:dimming}}
\end{figure}

\placefigure{fig:dimming}

Finally, we address the lack of a large-scale dimming signature in the EUV. 
Estimates of the mass associated with EUV dimmings showed it can represent a significant fraction of the total CME mass \citep[e.g.][]{2003A&A...400.1071H, 2004A&A...427..705Z}. If the CME originated from deep in the corona carrying some part of it, we would have expected to see a large EUV dimming commensurate to the amount of mass carried off by the CME.

The electron density, $n_e(r)$ drops dramatically with radial distance from the solar surface. The EUV intensities, proportional to $n^2_e(r)$, exhibit an even stronger fall-off. A small difference in initiation height of the ejected material can give a large difference in the signal of the coronal dimming. Conversely, the absence of a dimming suggests a large initiation height for our CME. To demonstrate this, we calculated proxies of the EUV intensities before, $I_{\mbox{\tiny{BG}}}$, and during, $I_{\mbox{\tiny{CME}}}$, the eruption of a hypothetical CME with its source region located on disk center. 
The background intensity per pixel is defined as
\begin{equation}
I_{\mbox{\tiny{BG}}} \equiv \int^{\infty}_{1\,\mbox{\tiny{R}}_\odot}{n^2_e(r) dr} \approx  \int^{10\,\mbox{\tiny{R}}_\odot}_{1\,\mbox{\tiny{R}}_\odot}{n^2_e(r) dr},
\end{equation}
where $r$ is distance from sun center. 
Similarly, the EUV intensity corresponding to the portion of the corona that is removed by the CME corresponds to 
\begin{equation}
I_{\mbox{\tiny{CME}}}  \approx  \int^{10\,\mbox{\tiny{R}}_\odot}_{r_0}{n^2_e(r) dr},
\end{equation}
assuming a starting height of $r_0$ for the CME.   A ratio $I_{\mbox{\tiny{CME}}}/I_{\mbox{\tiny{BG}}}$ significantly below one implies an intensity depletion caused by the removal of part of the corona by the CME. If this depletion occurs simultaneously over a large set of continuous observational pixels then we observe a  ÔdimmingÕ.  

Figure~\ref{fig:dimming} shows this ratio for starting heights $r_0$ in the range 1 -  2\,R$_{\odot}$.  For $n_e(r)$, we used the density profiles of  Baumbach-Allen \citep{2005psci.book.....A,1999JGR...104.9691G,1996ApJ...458..817G}. The first profile (solid line) corresponds to an `average' corona, whereas the other two correspond to coronal streamers. As expected, the dimming ratio $I_{\mbox{\tiny{CME}}}/I_{\mbox{\tiny{BG}}}$  is appreciably smaller than one only for starting heights below 1.4\,R$_{\odot}$. For CMEs starting higher up in the corona, this ratio is approximately one, meaning that for those CMEs no observable coronal dimming can be expected. Conversely, an absence of dimming implies that most of the plasma that contributed to the CME mass originated from above 1.4 R$_\odot$. For reference, we note that this height corresponds to the inner edge of the COR1 FOV. This limit is quite robust given that the dimming curves have little dependence on the employed density profile as can be seen in Figure \ref{fig:dimming}. 

Including temperature effects (through temperature response functions $R(T)$) into the determination of the EUV intensities would further compress the radial distance range of sensitivity of $I_{\mbox{\tiny{CME}}}/I_{\mbox{\tiny{BG}}}$. This is because for a multi-thermal line of sight only points with temperatures around the peak of the $R(T)$ of the employed channel  will contribute to the observed intensities. Moreover, EUV and SXR observations show small or little variations of the electron temperature in streamers \citep[e.g.][]{2002A&A...381.1049F,2002ApJ...571..999W}.   Finally note that a large-scale dimming, if present, would have manifested at least in one or more of the four EUVI channels. The response functions of these channels peak at  0.08, 0.90, 1.50 and 2.00 MK, which spans the bulk of the quiet sun temperature domain \citep[e.g.][]{1996ApJS..106..143B}. We found no evidence of significant flaring, which means that no large amount of plasma was heated off quiet sun conditions.

\subsection{Scenario of a large flux-rope}
To summarize, we offer the following scenario based on the above discussion. The event occurs over a largely empty filament channel, hence no or weak emission in 304 \AA\, or H$\alpha$, on the quiet sun. The flux-rope, that eventually forms the CME core, is visible for at least two days prior to the eruption in emission in the coronal lines (171 and 195~\AA). We deduce that the flux-rope mostly consists of hot material of about 1 MK and the material is concentrated at the bottom of the feature. The flux-rope is also situated at 0.15\,R$_\odot$ (bottom) above the surface, which is unusually high. We illustrated the large height of the overlying loop system, by comparing the cavity beneath it with the smaller cavity that is visible in EUVI B, and found that our cavity is 2.2 times larger. This is further corroborated by the magnetic field properties of the postulated source region. It shows that most of the overlying field lines have widely separated footpoints, comparable to the widths derived in \S~\ref{sec:source}. We believe that this is the first EUV observation of such a high-lying flux-rope. This observation was possible thanks to the large EUVI FOV and the wavelet processing. The EUVI A time series show activity at the flux-rope such as structures rising up and into the feature causing it to rotate. This is a process very suggestive of tether-cutting and explains the existence of hot plasma in the flux-rope. We think that the flux-rope is likely to have formed over several days via successive `tether-cutting'-like events, storing free magnetic energy that was later used to drive the CME \citep[e.g.][]{2006ApJ...641..590G}. The last and most severe one must have occurred on May 31st after which the flux-rope is clearly seen rising and the CME is in process. The scenario of the large flux-rope is consistent with the lack of heating or dimming signatures and the lack of any appreciable eruptive filament but is also consistent with the standard reconnection model of CMEs \citep[e.g.][]{2000JGR...10523153F}.

\section{Conclusion and implications}\label{sec_concl}
To conclude, we find that the CME erupted from the quiet sun along a polarity inversion line. The very low CME speed ($<300$ km/s) is in concert with the weak photospheric field ($<3$G). We attribute the lack of any significant low coronal signatures (flaring, EUV dimming, prominence disappearance) to the unusually large height from which the flux-rope lifted off. The bottom of the flux-rope was situated at 1.15\,R$_\odot$ and the overlying loop system exceeds 1.4\,R$_\odot$ (both measured from sun center). This interpretation is supported by the CME mass measurements and estimates of the expected EUV dimming intensity. 

Overall, this event is a typical streamer blowout CME in terms of its evolution and physical parameters. However, the multi-viewpoint analysis of this observation results in some very important implications:
\begin{enumerate}

\item We have unambiguously shown that large CMEs are not necessarily associated with clear low coronal or chromospheric features. Their disk-signatures may be weak or undetectable. Therefore, the lack of an obvious on-disk signature does not imply that a (partial) halo-CME is backsided, as has been assumed in numerous studies. The use of on-disk EUV or H$\alpha$ imaging as proxies of CMEs in the low corona cannot be considered as fully reliable for operational purposes. 

\item A CME can erupt from the quiet Sun where the field is weak. Our observation shows that no large filament or active region needs to be present in the pre-CME corona in order to initiate an eruption. The magnetic field configuration itself is more important than the plasma for studying CME initiation. Correlations of CMEs with prominences and flares will therefore vary depending on what instruments are used. Imaging instruments can only show structures that contain enough plasma (at the `right' temperature), for example active regions and their loops, but they are unable to track tenuous features like filament channels. Vector magnetograms, preferably in the upper chromosphere or corona,  could be used to detect magnetic configurations that can drive CMEs. 

\item This event is a good example of a `problem storm'. If this CME had been directed at Earth, but with southward Bz, it could have generated a moderate geomagnetic storm. Previous studies have shown that a significant fraction ($>10\%$) of geomagnetic storms could not be traced back to solar surface activity (even though a MC was observed) and were therefore called problem storms. Because these events are hard to observe face-on they are a challenge for space weather forecasters. During periods of high activity, these events may go unnoticed. However, these CMEs can interact with other interplanetary ejecta altering their properties and consequently their geoeffectiveness. Their detection is thus important.

\item This observation could not have been made without the dual STEREO observation. Our analysis thus strongly suggests that reliable prediction of Earth-directed CMEs can only be made by remote-sensing platforms away from the Sun-Earth line. 

\item Our observations present a new set of elements, which CME initiation models should be able to reproduce: the presence of a cavity prior to eruption (indicates a pre-existing flux-rope), the lack of any flaring, no dimming, the slow  evolution and the lack of cold prominence material. Further, the cavity indicates that the quiescent corona can be highly non-potential, which in turn implies the presence of free magnetic energy that may drive the CME.

\item We have shown in \S~\ref{sec:dimming} that the absence of an EUV dimming implies a large initiation height for CMEs. This observation shows that such CMEs can reach average CME masses (in our case $3\times10^{15}$g). To reach such a mass, the source region size of these CMEs should be quite extended as we found in  \S~\ref{sec:source}.
\end{enumerate}

Preliminary inspection of STEREO data indicates that events like this one are not rare. We speculate that this is a characteristic of the ongoing period of deep minimum. As a next step, we plan to analyze these events and establish their occurence rates, physical properties, possible effects on other CMEs during their interplanetary propagation and assess their geoeffective potential.

\acknowledgments The SECCHI data is produced by an international consortium of the NRL, LMSAL and NASA GSFC (USA), RAL and U. Bham (UK), MPS (Germany), CSL (Belgium), IOTA and IAS (France). 
We thank Yi-Ming Wang for producing the photospheric field map, Guillermo Stenborg for applying the wavelet technique and Yan Li for interpreting the STEREO-B in situ data. We benefitted from a useful discussion on filament channels with Sara Martin and Olga Panasenco.

\bibliographystyle{apj}
\bibliography{ms}


\end{document}